\documentclass[aps,prl,twocolumn,showpacs,superscriptaddress]{revtex4-1} 
\usepackage{graphicx}
\usepackage{subfigure}
\usepackage{bm}
\usepackage{latexsym,amsthm,amssymb}
\usepackage{booktabs,threeparttable}
\usepackage{makeidx}
\usepackage{amsmath}
\usepackage{amsfonts}
\usepackage{amssymb}
\usepackage{setspace}
\usepackage{slashed}
\usepackage{amsfonts}
\usepackage{amsthm}
\usepackage{amssymb}
\usepackage{amsmath}    
\usepackage{graphicx,floatrow}   
\usepackage{verbatim}   
\usepackage{color}      
\usepackage{subfigure}  
\usepackage{epsfig}
\usepackage{subfigure}
\usepackage{float}
\usepackage{epstopdf}
\usepackage{cases}
\usepackage{appendix}
\usepackage{bm}
\usepackage{lineno}
\newcommand{\norm}[1]{\left\Vert#1\right\Vert}
\usepackage[colorlinks=true, pdfborder=001, linkcolor=blue, anchorcolor=blue, citecolor=blue, urlcolor=blue]{hyperref}

\begin{document}
\title{Stable Black Hole with Yang-Mills Hair}
\author{Yuewen Chen}
\email{yuewen\_chern@amss.ac.cn}
\affiliation{Yau Mathematical Sciences Center, Tsinghua University, Beijing 100084, China}
\author{ Jie Du}
\email{jdu@tsinghua.edu.cn}
\affiliation{Yau Mathematical Sciences Center, Tsinghua University, Beijing 100084, China}
\affiliation{Yanqi Lake Beijing Institute of Mathematical Sciences and Applications, Beijing 101408, China}

\author{Shing-Tung Yau}
\email{Corresponding author: yau@math.harvard.edu}
\affiliation{Yau Mathematical Sciences Center, Tsinghua University, Beijing 100084, China}
\affiliation{Yanqi Lake Beijing Institute of Mathematical Sciences and Applications, Beijing 101408, China}
\affiliation{ Department of Mathematics, Harvard University, Cambridge, MA 02138, USA }


\begin{abstract}
We present  stable solution of static spherically symmetric Einstein-Yang-Mills equations with  the $SU(2) $ gauge group.
This solution is asymptotically flat and regular at $r = 0$  and  with nontrivial Yang-Mills(YM) connection.  With quantized values of the Arnowitt-Deser-Misner (ADM) mass, the solutions asymptotically approach the Schwarzschild solution and have zero global YM charges. 
Numerical evidences suggest that this solution is both linearly and nonlinearly stable and has a ring of generic
curvature singularities along the  horizon.  An effective counterexample to the no-hair conjecture is provided by this stable solution. Moreover,
the stable black hole solution suggests that the coupling of gauge field to gravity in early Universe 
will generate a new type  of black holes. Their stability means that these might be a possible new source of  primordial black holes left over from the early Universe and serves as a possible new candidate for dark matter.

\end{abstract}
\maketitle  
\textit{Introduction.}---In this letter, we are interested in the Einstein-Yang-Mills (EYM) equations. In 1988, a countable family of nontrivial static globally regular (i.e., nonsingular and asymptotically flat) solutions of spherically symmetric $SU(2)$ EYM equations was discovered by Bartnik and McKinnon numerically \cite{Bartnik}. After this pioneering work, a series of soliton and black hole solutions of spherically symmetric EYM equations were found \cite{2, 3, Kun}. Numerical solutions to EYM equations were also studied by many researchers \cite{cho1,cho2,cho3,cho4,nu1,nu2,nu3}. The critical behavior of spherically symmetric collapse of EYM equations were studied in \cite{cho1,cho2,cho3,cho4}. On the other hand, the major developments in the theoretical analysis were carried out by J. Smoller and his collaborators in a series of papers \cite{s1,s2,s3,s4,s5}. Smoller and Yau {\it et.al}  proved rigorously the existence of a globally defined smooth static solution \cite{s2} and showed that the $SU(2)$ EYM equations admit an infinite family of black-hole solutions with a regular event horizon \cite{s4}.
Meanwhile, they proved that there exist infinitely many smooth static regular solutions of EYM equations \cite{s1}.  However, Straumann and Zhou \cite{un1} demonstrated that Bartnik-McKinnon's solutions are unstable under linear perturbation. The colored black hole found numerically by Bizon is also  unstable\cite{un2,un3}. The nonlinear stability was studied in \cite{un4} and the work provided numerical evidence for the instability of the colored black hole solutions. This instability was also studied by Choptuik \cite{cho1}. The unstable mode for certain solution were also discovered in \cite{cho4,cho2}. 

 Based on these findings, there has been a common belief that the Einstein-Yang-Mills equations do not admit stable black hole solutions. 
However, all the solutions considered so far are defined in the classical solution space.
In this letter we shall show that, provided we enlarge the function space on which the field equations are defined to be of
 bounded variation($BV$) type,  stable YM black hole solutions indeed exist. 
  
The stable bounded variation solution will be constructed by solving the spherically symmetric $SU(2)$ EYM equation numerically. A six order weighted essentially nonoscillatory (WENO) scheme is adopted. Numerical experiments confirm the sixth-order convergence rate of our numerical scheme. An independent check work is also implemented by using a Discontinuous Galerkin (DG) method and the results are consistent with the results obtained by using WENO. We show that this solution are stable under the linear and nonlinear perturbation.

We should point out that the "no-hair" conjecture for YM black holes is called into question by our discovery. The structure of a stationary black hole is completely determined by global charges defined at spatial infinity such as Arnowitt Deser-Misner (ADM) mass, angular momentum, or electric charge, according to this conjecture.

Because the global YM charges are always zero for our stable EYM black holes, the ADM mass is still the only global parameter used to describe these solutions. Since the YM hair is not connected to any global charge that would prevent it from radiating away to infinity, the existence of such black holes is inconsistent with the fundamental tenet of the no-hair conjecture. Our  result complement the work of Bartnik-McKinnon  \cite{Bartnik}and Bizon \cite{2}.

 
\textit{Einstein-Yang-Mills equations and boundary conditions.}---The static, spherically symmetric  metric can be written as \cite{cho2}
$$ g=-A e^{-2 \delta} dt^2 +\frac{dr^2}{A} +r^2 (d\theta^2 +\sin^2\theta d\phi ^2).$$
Let $\tau_1,\tau_2,\tau_3$ denote the  Pauli matrices. The spherically symmetric Yang-Mills connection with $SU(2)$ gauge group can be written in the following  form
$$\mathfrak{A}=u \tau_3 dt+(W \tau_1 +\tilde{W} \tau_2) d\theta +(\cot \theta \tau_3+W \tau_2 -\tilde{W}\tau_1)\sin \theta d\phi ,$$
where $u$, $W$ and $\tilde{W}$ are functions of $t$ and $r$. We can choose $u=0$ and $\tilde{W}=0 $. Then the Yang-Mills field corresponding to the simplified gauge potential becomes
$$\mathfrak{A}=W \tau_1  d\theta +(\cot \theta \tau_3+W \tau_2 )\sin\theta d\phi .$$
The Yang-Mills curvature tensor is given by
$$F=W' \tau_1 dr \wedge d\theta+W' \tau_2  dr\wedge \sin \theta d \phi -(1-W^2) \tau_3 d \theta \wedge \sin \theta d \phi.$$
The radial and angular magnetic curvatures are  defined as 
$$B_L=-\frac{1-W^2}{r^2}\tau_3, B_T=\frac{W'}{r}\tau_1.$$
The Einstein-Yang-Mills equations with $SU(2)$ gauge potential have been derived in many papers. One can find  more details in \cite{s2,Bartnik,cho2}. For the sake of brevity, we just write down the Einstein-Yang-Mills system  directly,
\begin{align}
	r^2 AW^{''} &=(\frac{(W^2-1)^2}{r}+r(A-1))W^{'}+W(W^2-1)  \label{eq:4},\\
	r A^{'} &=1-\frac{(W^2-1)^2}{r^2}-A(2{W^{'}}^2+1)  \label{eq:5}, \\
	\delta^{'}&=-\frac{2(W^{'})^2}{r}\label{eq:3}
\end{align}
with  boundary conditions
\begin{align}
	W(0) &=\pm 1,\\
	W(\infty) &=\mp 1,\\
	A(0) &=1,\\
	\delta(0) &=0,
	\end{align}
where ${}^{'}=\frac{\partial }{\partial r}$.
Note that (\ref{eq:4}) and (\ref{eq:5}) do not involve $\delta$, so one can first solve these two equations for $A$ and $W$ and then use (\ref{eq:3}) to obtain $\delta$.
In this paper, we need a new coordinate transformation
\begin{align}
	x&=f(r)=\ln(r) \label{eq:t},\\
	r&=e^{x},
\end{align}
which gives 
\begin{align*}
	\frac{d}{dr}&=\frac{1}{r}\frac{d}{dx},~~\frac{d^2}{dr^2}=-\frac{1}{r^2}\frac{d}{dx}+\frac{1}{r^2}\frac{d^2}{dx^2}.
\end{align*}
Then the equations for $W$ and $A$ become
\begin{align}
	&A W_{xx}-(2A-1+\frac{(W^2-1)^2}{r^2}) W_x +(1-W^2)W =0 ,\label{eq:7}\\
	&-(1+\frac{2}{r^2}W_x^2) A+1-\frac{1}{r^2}(W^2-1)^2 =A_x.\label{eq:8}
\end{align}
We consider the following evolution version of (\ref{eq:7}) and (\ref{eq:8})
\begin{align}
	&A W_{xx}-(2A-1+\frac{(W^2-1)^2}{r^2}) W_x +(1-W^2)W =W_t ,\label{eq:9}\\
	&-(1+\frac{2}{r^2} W_x^2) A+1-\frac{1}{r^2}(W^2-1)^2 =A_x ,\label{eq:10}
\end{align}
and aim to compute this system to steady state numerically. 


\indent
\textit{Numerical scheme.}---One of the most important features of system (\ref{eq:7})-(\ref{eq:8}) is the degeneration. The minimum of $A$ is very small such that there is a sharp front in $W$. There are many numerical schemes designed for solving 
degenerate convection-diffusion equations. A high order finite difference WENO scheme for nonlinear degenerate parabolic equations was first developed by Liu, Shu and Zhang \cite{shu1}. 
In this section, we will design a sixth order WENO scheme for solving (\ref{eq:9})-(\ref{eq:10}). For more  numerical detail, one can see Chapter 2 of supplemental material.
The computational domain is $[-5,5]$. The boundary condition is given by
$$W(-5)=-1,~W(5)=1,~A(-5)=1.$$ 
The initial condition is constructed as follows:
\begin{align}
	W(x)=\tanh(40(x+0.5)).
\end{align}
The initial condition of $A $ can be solved from (\ref{eq:8}) with $A(-5)=1$. 

\begin{figure*}[htbp]
\centering
\subfigure[]{
\begin{minipage}[b]{0.48\linewidth}
\includegraphics[width=1.1\linewidth]{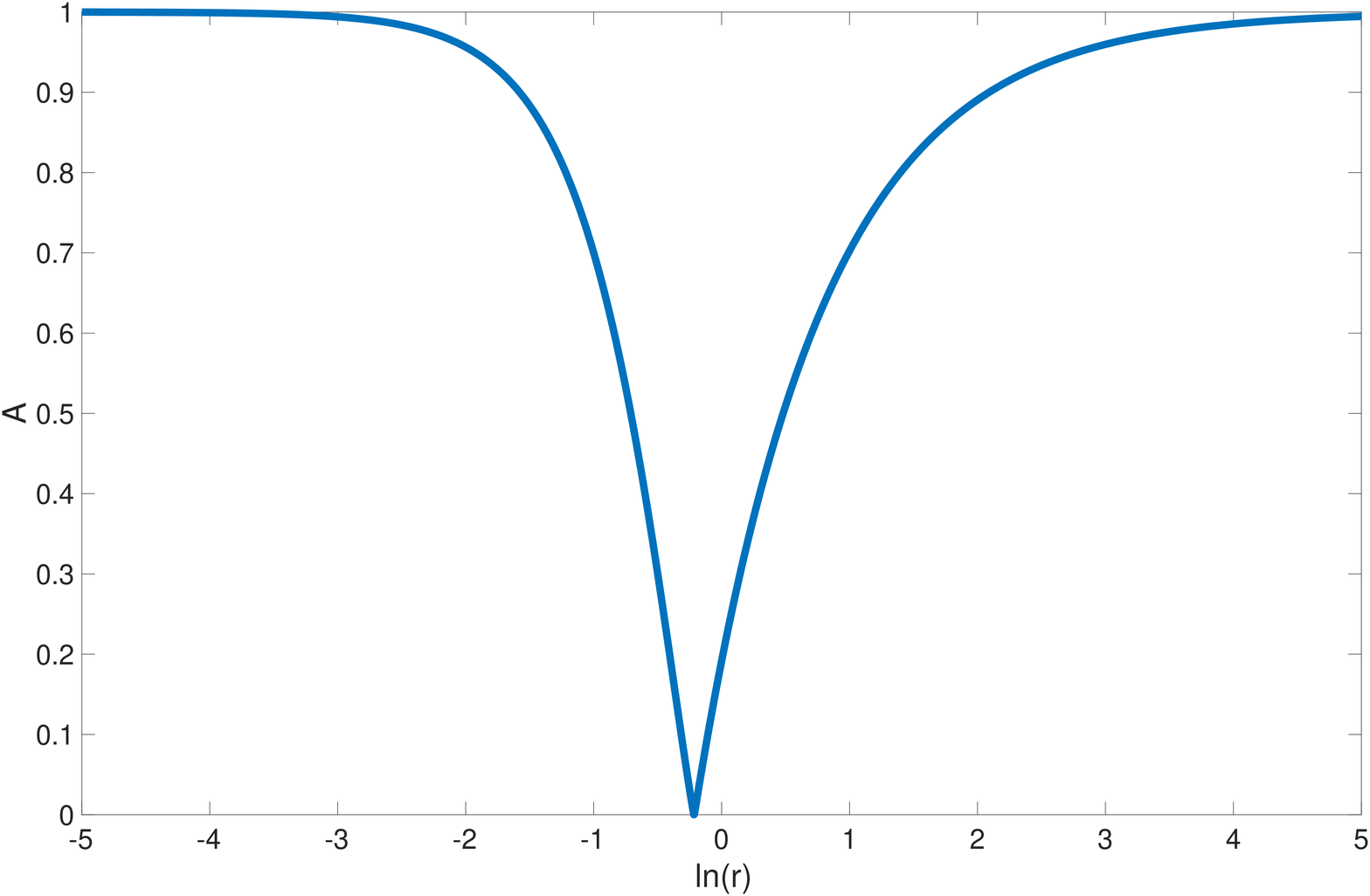}
\end{minipage}}
\subfigure[]{
\begin{minipage}[b]{0.48\linewidth}
\includegraphics[width=1.1\linewidth]{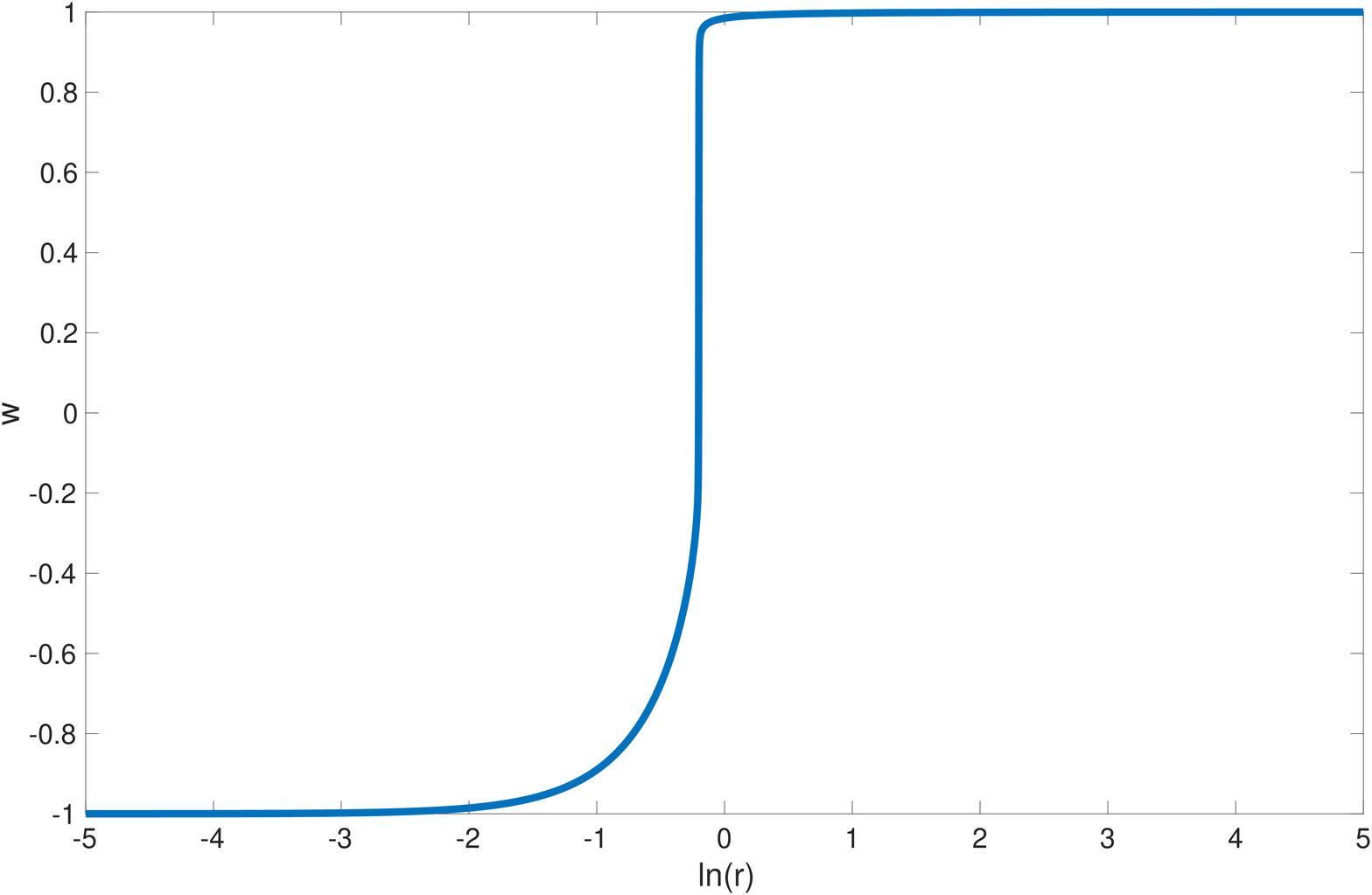}
\end{minipage}}
\subfigure[]{
\begin{minipage}[b]{0.48\linewidth}
\includegraphics[width=1.1\linewidth]{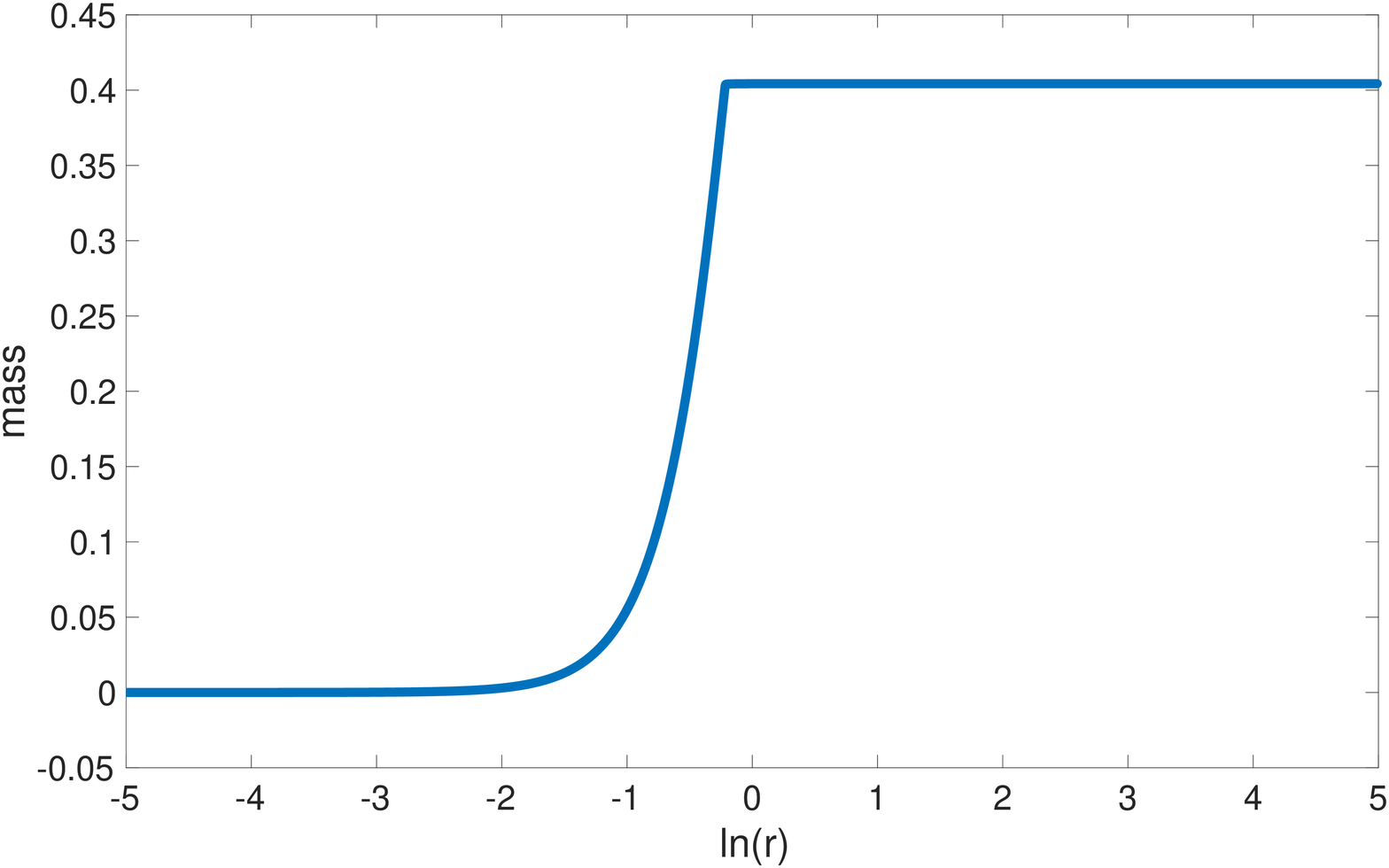}
\end{minipage}}
\subfigure[]{
\begin{minipage}[b]{0.46\linewidth}
\includegraphics[width=1.12\linewidth]{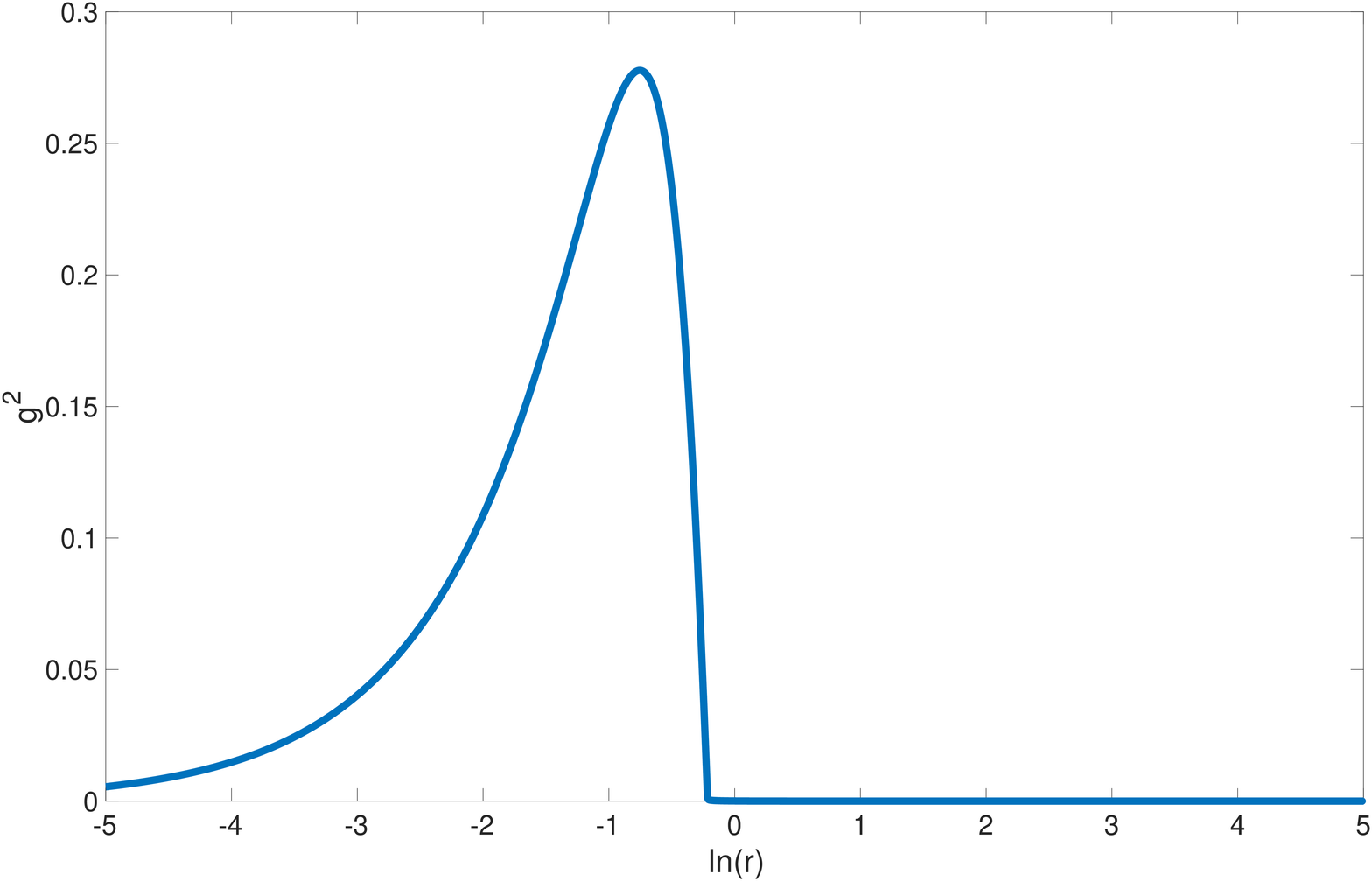}
\end{minipage}}
\caption{ (a): The $A$ is regular $r=0$ . the horizon at $x_o=-0.21$,$A(x_o)=0$. (b) :$W$  approximate $1$ in the far field and close to -1  near $r=0.$
(c): The mass  increase to the total mass $M=0.4.$ (d) :The effective charge $g^2$ close to 0.3 near the horizon and drop to 0 in the exterior .
} \label{fig1}
\end{figure*}
We compute the $L^\infty$ and $L_1$ errors and orders in Table \ref{table1}. Since we do not know the exact solution of EYM equations, as a substitution we use numerical solutions obtained on a fine grid with $2^{15}$ points as approximations to the exact solutions, which are denoted by $W_{15}$ and $A_{15}$. 
This table show that our numerical scheme converge to the right solution with sixth-order.  
The mass is defined by $$m=\frac{r}{2}(1-A).$$ We plot the mass in Fig.~\ref{fig1}.

The mass function $m(r)$ increase to the total mass $M=0.40.$ Bizon's  solution is quite similar to the Resissner-Nordstrom solution near to the event horizon\cite{2}. In order to compare with Bizon's cloled black hole, we define the RN (magnetic) charge as Bartink $$g^2(r)=2r (M-m). $$ 
Fig.~\ref{fig1} indicates that $g^2$ is approximately 0.3 near the horizon. However, Bizon's colored black hole have the charge $g^2 \approx 1$ close to horizon\cite{2}.  In the far-field region, $W \to 1$ and the YM charge disappear. 
The ADM mass is the sole global parameter used to describe the metric at infinity. The Schwarzschild solution must therefore be the only black hole solution, in accordance with the no-hair conjecture.

\begin{table}[!htp]
	\resizebox{9cm}{1.1cm}{\begin{tabular}{|c|cccc|cccc|}
		\hline
		N       &$ \norm{W_{15}-W_h}_{L^1}$    & order   &$ \norm{W_{15}-W_h}_\infty$   & order&$ \norm{A_{15}-A_h}_{L^1}$    & order   &$ \norm{A_{15}-A_h}_\infty$   & order\\\hline
		$2^9$           &   3.55E-09  &   --    &     5.50E-08   &  --   &      8.73E-04 & --     &  1.71E-03&  --   \\
		$2^{10} $      &   2.28E-10  &    3.96&     3.94E-09  &  3.80&   6.06E-05   &3.85   &  1.21E-04& 3.83\\
		$2^{11}$        &   4.78E-12  &    5.58&     8.99E-11   &  5.45&   1.23E-06   & 5.62  & 2.45E-06& 5.63\\
		$ 2^{12} $     &    1.00E-13  &    5.62&     1.91E-12   &   5.55&   2.46E-08  & 5.65  & 4.90E-08&5.65 \\
		$ 2^{13}$      &    1.50E-15  &     6.05&    2.90E-14  &   6.00&   3.76E-10   &  6.02 & 7.60E-10&6.01  \\
		\hline	
	\end{tabular}}
	\caption{\label{table1} Error table for sixth order  WENO scheme in the smooth region of the solution.}
\end{table}
\indent
\textit{Linear stability analysis of the static solutions.}---In this section, we analyses the linear stability of  our solution, more precisely, we just  consider the mode stability at this section. Our work provides  strong numerical evidence  for the linear stability, however the numerical evidence can't take the  place of the strict proof,a rigorous analytical proof will be given will be given in our next work. The linear stability problem can be reduced to an ODE eigenvalue problem. 
For a small amplitude of perturbation departures from our static solution, denoted by $W_0$ and $A_0$, we make the following perturbative ansatz:
\begin{align}
	W(t,r)&=W_0(r)+\varepsilon \tilde{W}(t,r), \\
	A(t,r)&=A_0(r)(1+\varepsilon \tilde{A}(t,r)), \\
	\delta(t,r) &=\delta_0(r)+\varepsilon\tilde{\delta}(t,r).
\end{align}

Separation of variables 
\begin{align}
	\tilde{W}(t,r)&=\xi(r)e^{i \sigma t}, \label{eq:linear1}\\
	\tilde{A}(t,r)&=\alpha(r)e^{i\sigma t}, \label{eq:linear2}\\
	\tilde{\delta}(t,r) &=\beta(r) e^{i \sigma t}. \label{eq:linear3}	
\end{align}

We  reduce  the following ODE eigenvalue problem 

\begin{align}
p(x)\xi\sigma^2&=-\xi_{xx}g_1(x)^2+\Omega_1(x) \xi_x+\Omega_2(x) \xi  \label{eq:6.21},\\
	\xi(-5) &=0,~\xi(5) =0.
\end{align}

where $p(x),\Omega_1,\Omega_2,g_1$ are defined in supplemental material.	
We use the second order  center finite difference scheme to approximate (\ref{eq:6.21}):
\begin{align}
	p_j \xi_j \sigma^2 =-\frac{g_1(x_j)^2}{h^2}(\xi_{j+1}-2\xi_j+\xi_{j-1})+  \nonumber \\
	\Omega_1(x_j)\frac{1}{2h}(\xi_{j+1}-\xi_{j-1})+\Omega_2(x_j)\xi_j \label{eq:6.24}
\end{align}
for $j=1,..n$ with the following boundary conditions 
\begin{align}
	\xi_{0} &=0,\\
	\xi_{n+1} &=0.
\end{align}
Then we obtain an algebraic eigenvalue problem which can be written as
\begin{align}
	\Lambda \Xi = \sigma ^2P  \Xi,
\end{align}
where  $\Lambda$ is a tridiagonal matrix,  $\Xi$ is the vector with components $\xi_j$ and $P$ is  a diagonal matrix whose $j$th element is $p_j$. We now find the eigenvalues by using a standard matrix technique such as the Rayleigh quotient iteration and QR Algorithm.
The  first eigenvalue with  is
$$\min(\sigma^2)=462.04855,$$
which means $\sigma$ is a real number. Substituting this value into (\ref{eq:linear1})-(\ref{eq:linear3}), we can see that the amplitude of the perturbation will not grow. But we still need a strong evidence that the perturbation will decay. Hence, we consider the nonlinear perturbation of the static solution in the next section.  

\textit{Nonlinear perturbation.}---The EYM equation can be written as 
\begin{align}
	r^2e^{\delta}(\frac{e^\delta}{A}W_t)_t &=AW_{xx}+(1-2A-\frac{(1-W^2)^2}{r^2})W_x\nonumber\\
	&~~~+(1-W^2)W,\\
	A_x &=-(1+\Pi^2 +\Phi^2)A+1-\frac{(1-W^2)^2}{r^2}, \\
	\delta_x &=-(\Pi^2+\Phi^2),
\end{align}
where 
\begin{align}
	\Pi &=\frac{e^\delta}{A} W_t,~
	\Phi =\frac{\sqrt{2}W_x}{r}.
\end{align}
We give the initial perturbations as:
\begin{align}
	 \text{data 1}: W(0,x) &=W_0(x)+0.1e^{-50(x-3)^2}, W_t(0,x) =0,\\
	\text{ data 2}: W(0,x) &=W_0(x)+0.1e^{-50(x+3)^2},W_t(0,x) =0. 
\end{align}
where $W_0(x)$ is static solution.
We introduce auxiliary variables $\alpha$ and $\beta$ defined as below and thus can rewrite the EYM equations as
\begin{align}
		c&=2A-1+\frac{(W^2-1)^2}{r^2}, \\
	\alpha &=\frac{e^{2\delta}r^2}{A}, \\
	\beta  & =r^2 e^{\delta}(\frac{e^\delta}{A} )_t,\\
	\alpha  W_{tt}+\beta W_t &=AW_{xx}-cW_x+(1-W^2)W,\\
	A_x &=-(1+\Pi^2 +\Phi^2)A+1-\frac{(1-W^2)^2}{r^2}, \\
	\delta_x &=-(\Pi^2+\Phi^2).
\end{align}
One can find the numerical detail in Chapter 4 of supplemental material.
 Next, we define $\Delta W(t,x)=W(t,x)-W_0(x)$ and compute the evolution of $\Delta W(t,x)$. \\ Based on the numerical evidence,
 \textit{we speculate that the static solution $W(x,t)$ is  stable in the $BV$ norm. More precisely, we conjecture that $\|\Delta W\|_{BV} \to 0$ as $t \to \infty$.}
 
\begin{figure*}[htbp]
	\begin{minipage}[b]{1\linewidth}
		\centering
	\includegraphics[width=18cm]{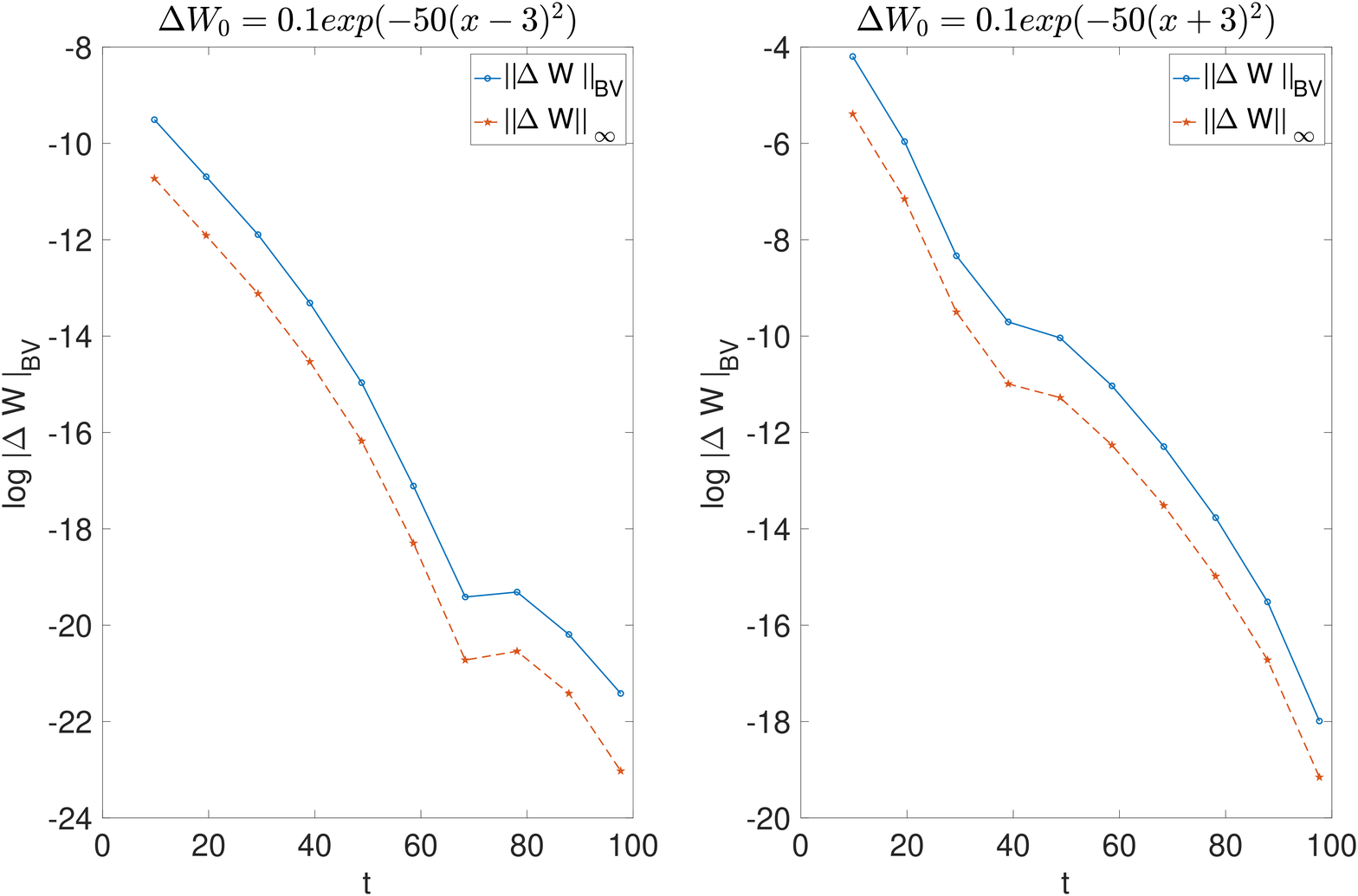}\\
	\caption{Left:for the perturbation Data 1, the $\log (\|\Delta W\|_{BV})$ and $\log (\|\Delta W\|_{L^\infty})$  decay to $-\infty$  as  $t \to \infty$. Right:for the perturbation Data 2, the $\log (\|\Delta W\|_{BV})$ and $\log (\|\Delta W\|_{L^\infty})$  decay to $-\infty$   as  $t \to \infty$,which mean that static solution is stable.}
	\label{fig50}	
\end{minipage}
           \end{figure*}

 As shown in Fig.\ref{fig50}, both the $L^{\infty}$ norm and the $BV$ norm of $\Delta W$ decay to 0 as the time increases. Therefore, the perturbation solution converges to the steady state solution $W_0$, which indicates that our solution $W_0$ is nonlinearly stable.

~~

\textit{Geometry and light cones in EYM spacetime }---
The   Riemann curvature of EYM  spacetime are regular at $r=0$ (i.e.,  $x=-\infty$), but are singular at $x=-0.21$,furthermore, the Kretschmann scalar and Ricci scalar also blow up at the horizon but keep regular at $r=0$,which  are  different from Schwarzschild spacetime. The light cone structure shows  that there is an apparent  horizon  at $x=-0.21$. For more detail, such as the null geodesics, the time-like geodesic and the space-time diagram in Eddington-Finkelstein coordinates.  One can find in  Chapter 5  of  supplemental material.

\textit{Conclusion and Discussion}---
In this paper, we  give strong evidence of stability of our solution. In the future, rigorous analytical proof will be given. We shall also study more general metric where the tensor will have more mixed terms. On the physical side, the stability of the solutions suggests that, in early universe when gauge field plays a dominant role in the evolution, the coupling of gravity and gauge field would generate a new type of stable black holes. These primordial black holes left over from early Universe
might be a source of dark matter at the  present epoch \cite{dm,bcm}.


\begin{thebibliography}{99}   
	\bibitem{Bartnik}R. Bartnik and J. McKinnon, Particle-like solutions of the Einstein Yang-Mills equations, \href{https://doi.org/10.1103/PhysRevLett.61.141}{Phys. Rev. Lett. \textbf{61}, 141 (1988)}. 	
	\bibitem{2}P. Bizon, Colored black holes, \href{https://doi.org/10.1103/PhysRevLett.64.2844}{Phys. Rev. Lett. \textbf{64}, 2844 (1990)}.
	\bibitem{3} M. S. Volkov and D. V. Gal'tsov, NonAbelian Einstein-Yang-Mills black holes, Sov. J. Nucl. Phys. \textbf{51}, 1171 (1990).
	\bibitem{Kun}  H. P. K\"unzle and A. K. M. Masood-ul-Alam, Spherically symmetric static $SU(2)$ Einstein-Yang-Mills fields,  \href{https://doi.org/10.1063/1.528773}{J. Math. Phys. \textbf{31},  928 (1990)}.
	\bibitem{cho1} M. W. Choptuik, J.  Chmaj, and P. Bizo\'n, Critical Behavior in Gravitational Collapse of a Yang-Mills Field,  \href{https://doi.org/10.1103/PhysRevLett.77.424}{Phys. Rev. Lett. \textbf{773},  424-427 (1996)}.
	\bibitem{cho2}M. Maliborski and O. Rinne, Critical phenomena in the general spherically symmetric Einstein-Yang-Mills system, \href{https://doi.org/10.1103/PhysRevD.97.044053}{Phys. Rev.  D \textbf{97}, 044053  (2018)}.
	\bibitem{cho3} M. W. Choptuik,  E. W. Hirschmann, and R. L. Marsa. New Critical Behavior in Einstein-Yang-Mills Collapse, \href{https://doi.org/10.1103/physrevd.60.124011}{Phys. Rev.  D \textbf{60},  124011 (1999)}.
	\bibitem{cho4} O. Rinne, Formation and decay of Einstein-Yang-Mills black holes,  \href{https://doi.org/10.1103/physrevd.90.124084}{Phys. Rev.  D \textbf{90}, 124084 (2014)}.
	\bibitem{nu1} A. Zengino$\mathrm{\breve{g}}$lu, A hyperboloidal study of tail decay rates for scalar and Yang-Mills fields,  \href{https://doi.org/10.1088/0264-9381/25/17/175013}{Class. Quantum Grav. \textbf{25}, 175013  (2008)}.
	\bibitem{nu2} M. P\"urrer  and P. C. Aichelburg, Tails for the Einstein-Yang-Mills system, \href{https://doi.org/10.1088/0264-9381/26/3/035004}{Class. Quantum Grav. \textbf{26},  035004 (2009)}.
	\bibitem{nu3}P. Bizo\'n, A. Rostworowski, and A. Zenginou$\mathrm{\breve{g}}$lu, Saddle-point dynamics of a Yang-Mills field on the exterior Schwarzschild spacetime, \href{https://doi.org/10.1088/0264-9381/27/17/175003}{Class. Quantum Grav. \textbf{27},  175003 (2010)}.
	\bibitem{s2} J. A. Smoller, A. G. Wasserman, S.-T. Yau, and J. B. McLeod, Smooth static solutions of the Einstein-Yang/Mills equation, \href{https://doi.org/10.1007/BF02100288}{Commun. Math. Phys. \textbf{143}, 115-147 (1991)}.
	\bibitem{s4} J. A. Smoller, A. G. Wasserman, and S.-T. Yau, Existence of black-hole solutions for the Einstein-Yang/Mills equations, \href{https://doi.org/10.1007/BF02097002}
{Commun. Math. Phys. \textbf{154}, 377-401 (1993)}.
	\bibitem{s1}J. A. Smoller, A. G. Wasserman, Existence of infinitely-many smooth, static, global solutions of the Einstein/Yang-Mills equations, \href{https://doi.org/10.1007/BF02096771}{Commun. Math. Phys. \textbf{151}, 303-325 (1993)}.
	\bibitem{s3} J. A. Smoller and A. G. Wasserman,  Reissner-Nordstr\"om-like solutions of the spherically symmetric $SU(2)$ Einstein/Yang-Mills equations,  \href{https://doi.org/10.1063/1.532224}{J. Math. Phys. \textbf{38}, 6522-6559 (1997)}.
	
	\bibitem{s5} J. A. Smoller and A. Wasserman, Regular solutions of the Einstein-Yang-Mills equations, \href{https://doi.org/10.1063/1.530963}{J. Math. Phys. \textbf{36}, 4301-4323 (1995).}
	\bibitem{un1}N. Straumann and Z. H. Zhou, Instability of the Bartnik-mckinnon solution of the Einstein-Yang-Mills Equations, \href{https://doi.org/10.1016/0370-2693(90)91188-H} {Phys. Lett. B \textbf{237}, 353-356 (1990)}.
	\bibitem{un2}N. Straumann and Z. H. Zhou, Instability of a colored black hole solution, \href{https://doi.org/10.1016/0370-2693(90)90951-2}{Phys. Lett. B \textbf{243}, 33-35 (1990)}.
	\bibitem{un3} P. Bizon and R. M. Wald, The $n=1$ colored black hole is unstable, \href{https://doi.org/https://doi.org/10.1016/0370-2693(91)91243-O}{Phys. Lett. B \textbf{267}, 173-174 (1991)}.
	\bibitem{un4}Z. H. Zhou and N. Straumann, Nonlinear perturbations of Einstein-Yang-Mills solitons and non-abelian black holes, \href{https://doi.org/10.1016/0550-3213(91)90439-5}{Nucl. Phys. B \textbf{360}, 180-196 (1991)}.
	\bibitem{shu1} Y.  Liu,  C.-W Shu, and M.  Zhang, High order finite difference WENO schemes for nonlinear degenerate parabolic equations, \href{https://doi.org/10.1137/100791002}{SIAM J. Sci. Comput. \textbf{33} (2), 939–965 (2011).}
	\bibitem{shu2} C.-W. Shu,  Essentially non-oscillatory and weighted essentially non-oscillatory schemes for hyperbolic conservation laws, in Advanced Numerical Approximation of Nonlinear Hyperbolic Equations. B. Cockburn, C. Johnson, C.-W. Shu, and E. Tadmor (Editor: A. Quarteroni),  Lecture Notes in Mathematics,  volume \textbf{1697}, Springer, pp. 325-432 (1998).
	\bibitem{dm} B. Carr  and F. K\"uhnel,  Primordial Black Holes as Dark Matter: Recent Developments,  \href{https://doi.org/10.1146/annurev-nucl-050520-125911}{Annu. Rev. Nucl. Part. Sci. \textbf{70}, 355–394 (2020).}
	\bibitem{bcm} S. Bird,  I. Cholis, J. B. Mu\~noz,  Y. Ali-Ha\"{\i}moud,  M. Kamionkowski, E. D.  Kovetz, A. Raccanelli,  and A. G. Riess, Did LIGO Detect Dark Matter?  \href{https://link.aps.org/doi/10.1103/PhysRevLett.116.201301}{Phys. Rev. Lett. \textbf{116},  201301 (2016).}
  		                    
\end{thebibliography}
\end{document}